\documentclass[conference]{IEEEtran}
\IEEEoverridecommandlockouts
\usepackage[utf8]{inputenc}
\usepackage[T1]{fontenc}
\usepackage[english]{babel} 
\usepackage{cite}
\usepackage{amsmath,amssymb,amsfonts}
\usepackage{algorithmic}
\usepackage{graphicx}
\usepackage{textcomp}
\usepackage{xcolor}
\usepackage{booktabs} 
\usepackage{float}
\usepackage{url}
\usepackage{listings}
\usepackage{xcolor}

\def\BibTeX{{\rm B\kern-.05em{\sc i\kern-.025em b}\kern-.08em
    T\kern-.1667em\lower.7ex\hbox{E}\kern-.125emX}}

\begin{document}


\title{OkanNet: A Lightweight Deep Learning Architecture for Classification of Brain Tumor from MRI Images}

\author{\IEEEauthorblockN{Okan Uçar}
\IEEEauthorblockA{\textit{Graduate School of Natural and Applied Sciences} \\
\textit{Department of Computer Engineering}\\
\textit{Ege University}\\
Izmir, Turkey \\
okanucar2000@hotmail.com}
\and
\IEEEauthorblockN{Murat Kurt}
\IEEEauthorblockA{\textit{International Computer Institute} \\
\textit{Ege University}\\
Izmir, Turkey \\
murat.kurt@ege.edu.tr}
}

\maketitle

\begin{abstract}
Medical imaging techniques, especially Magnetic Resonance Imaging (MRI), are accepted as the gold standard in the diagnosis and treatment planning of neurological diseases. However, the manual analysis of MRI images is a time-consuming process for radiologists and is prone to human error due to fatigue. In this study, two different Deep Learning approaches were developed and analyzed comparatively for the automatic detection and classification of brain tumors (Glioma, Meningioma, Pituitary, and No Tumor). In the first approach, a custom Convolutional Neural Network (CNN) architecture named ``OkanNet,'' which has a low computational cost and fast training time, was designed from scratch. In the second approach, the Transfer Learning method was applied using the 50-layer ResNet-50~\cite{resnet_ref} architecture, pre-trained on the ImageNet dataset. In experiments conducted on an extended dataset compiled by Masoud Nickparvar containing a total of $7,023$ MRI images, the Transfer Learning-based ResNet-50 model exhibited superior classification performance, achieving $96.49\%$ Accuracy and $0.963$ Precision. In contrast, the custom OkanNet architecture reached an accuracy rate of $88.10\%$; however, it proved to be a strong alternative for mobile and embedded systems with limited computational power by yielding results approximately $3.2$ times faster ($311$ seconds) than ResNet-50 in terms of training time. This study demonstrates the trade-off between model depth and computational efficiency in medical image analysis through experimental data.
\end{abstract}


\label{Sec:Intro}
\section{Introduction}
Brain tumors are defined as abnormal and uncontrolled cell growths within the skull and constitute a significant portion of cancer-related deaths worldwide. The type, location, and stage of the tumor are of vital importance in determining the treatment protocol (surgical intervention, radiotherapy, or chemotherapy) to be applied. Today, Magnetic Resonance Imaging (MRI) technology is widely used in the imaging of brain tumors due to its lack of ionizing radiation and its ability to present soft tissue contrast with high resolution. However, the examination of hundreds of MRI slices belonging to a patient one by one by expert radiologists creates a serious workload and brings with it the risk of oversight. In this context, the development of Computer-Aided Diagnosis (CAD) systems has become a necessity to support clinical decision-making processes and shorten the diagnosis time.

In recent years, developments in Artificial Intelligence (AI) and especially Deep Learning have produced revolutionary results in medical image analysis~\cite{medical_dl}. Unlike traditional machine learning methods (SVM, Random Forest, etc.), Deep Learning-based Convolutional Neural Networks (CNN) have the ability to automatically learn complex features (edges, textures, shapes, gradients) in images without human intervention. This capability has made CNNs the most powerful tool in distinguishing heterogeneous lesions, such as tumors, from healthy tissue.

Studies on brain tumor classification in the literature are generally gathered around two main axes: Custom architectures designed from scratch and Transfer Learning methods. Networks designed from scratch can be optimized specifically for the problem space and can work with fewer parameters. However, for these networks to achieve high success rates, very large datasets and careful hyperparameter optimization are needed. On the other hand, the Transfer Learning method, which uses ready-made models (ResNet, VGG, Inception, EfficientNet, etc.) trained with millions of images on massive datasets like ImageNet, allows reaching very high success rates with a limited amount of medical data. Transfer learning is a critical strategy for increasing the generalization ability of the model, especially in the field of medical imaging where training data is limited.

In this study, a deep learning-based hybrid classification system is proposed using a comprehensive MRI dataset containing 4 different classes: Glioma, Meningioma, Pituitary Tumor, and Healthy Tissue. The main motivation of the study is not only to achieve a high accuracy rate but also to analyze the relationship between computational cost and performance. For this purpose, two different strategies were followed within the scope of the study:
\begin{enumerate}
    \item \textbf{Custom Architecture Design (OkanNet):} A special CNN architecture requiring fewer layers, capable of fast training, and demanding low hardware resources was designed and trained from scratch.
    \item \textbf{Transfer Learning (ResNet-50):} To utilize the power of deep architectures, the ResNet-50 model, famous for its residual blocks, was adapted to the problem (fine-tuning), and its performance was analyzed.
\end{enumerate}

The remainder of the study is organized as follows: Section II details the dataset used and the proposed methods, Section III presents the experimental findings and comparative analyses, and Section IV discusses the results obtained.

\label{Sec:appalgo}
\section{Applied Algorithms and Methods}

In this section, the characteristics of the dataset used in the study, data preprocessing steps, the mathematical infrastructure of the developed unique CNN architecture (OkanNet), and the ResNet-50 model used for Transfer Learning are detailed. Additionally, the hardware environment where experiments were performed and the training hyperparameters are presented.

\subsection{Dataset and Preprocessing}
Within the scope of the study, the ``Brain Tumor MRI Dataset'' compiled by Masoud Nickparvar~\cite{dataset_ref} and open to researchers was used\footnote{Dataset Access: \url{https://www.kaggle.com/datasets/masoudnickparvar/brain-tumor-mri-dataset}}. The dataset consists of a total of $7,023$ images obtained from T1, T2, and FLAIR weighted MRI sequences. The dataset is divided into four main classes: Glioma, Meningioma, Pituitary, and No Tumor.

The following preprocessing steps were applied to the raw data to increase the training stability of the models:
\begin{itemize}
    \item \textbf{Resizing:} All MRI images with different resolutions were reduced to the $224 \times 224$ pixel size, which is the input layer standard of the ResNet-50 architecture, using the bi-cubic interpolation method.
    \item \textbf{Channel Conversion:} MRI images can be single-channel (grayscale) by nature. Since deep learning models generally expect 3-channel (RGB) input, grayscale images were converted to 3-channel space (Grayscale-to-RGB).
    \item \textbf{Data Augmentation:} Random rotation, horizontal flipping, and translation operations were applied to the training data to prevent deep networks from memorizing (overfitting) and to artificially increase data diversity.
\end{itemize}

\subsection{Method 1: Custom CNN Architecture (OkanNet)}
Inspired by pioneering studies in the literature such as LeNet-5 and AlexNet~\cite{alexnet_ref}, a unique CNN architecture with a low parameter count but high feature extraction capacity, specific to the brain tumor classification problem, was designed. This model, named OkanNet, consists  basically of three consecutive Convolutional Blocks followed by Classification Layers.

\subsubsection{Convolution and Activation Layers}
In each block, filters (kernels) of size $3 \times 3$ were used to capture local features (edges, corners, textures) on the image. The convolution operation is mathematically expressed by Equation \ref{eq:conv}:

\begin{equation}
y(i,j) = \sum_{m} \sum_{n} x(i+m, j+n) \cdot w(m,n) + b
\label{eq:conv}
\end{equation}

Here, $x$ represents the input image, $w$ the filter weights, and $b$ the bias value. After each convolution operation, the ReLU (Rectified Linear Unit) activation function was used to provide non-linearity to the network and to suppress negative values:

\begin{equation}
f(x) = \max(0, x)
\end{equation}

By using 16, 32, and 64 filters respectively in the model, it was aimed to learn more abstract features as the network deepens.

\subsubsection{Pooling \& Dropout}
To reduce dimensions and lighten the computational load, Max Pooling with a size of $2 \times 2$ and a stride of $2$ was applied in each block. Additionally, a $50\%$ ``Dropout'' technique was used in the fully connected layers to prevent overfitting.

\subsection{Method 2: Transfer Learning of ResNet-50}
As Deep Neural Networks deepen, they become harder to train and reach performance saturation due to the ``Vanishing Gradient'' problem. The ResNet (Residual Network) architecture developed by He et al. solved this problem with the ``Skip Connections'' structure \cite{resnet_ref}.

In this study, the 50-layer ResNet-50 model, pre-trained with millions of images on the ImageNet dataset, was used. The ``Fine-Tuning'' method was adopted as the transfer learning strategy. The 1000-class layer at the end of the model was removed and replaced with a new fully connected layer with $4$ outputs (Glioma, Meningioma, Pituitary, No Tumor). The basic mathematical expression of the ResNet block is given in Equation \ref{eq:resnet}:

\begin{equation}
y = F(x, \{W_i\}) + x
\label{eq:resnet}
\end{equation}

Here, $x$ represents the data entering the block, $F(x)$ the learned residual function, and $y$ the output.

\subsection{Experimental Setup and Evaluation Metrics}
All experiments were carried out on NVIDIA GeForce RTX 2060 GPU hardware using MATLAB R2023b and Deep Learning Toolbox libraries. Stochastic Gradient Descent with momentum (SGDM) was preferred as the optimization algorithm throughout the training process.

\subsubsection{Hyperparameters}
For a fair comparison of the models, both architectures were trained with the following common parameters:
\begin{itemize}
    \item \textbf{Epoch Count:} 8
    \item \textbf{Mini-Batch Size:} 32
    \item \textbf{Learning Rate:} $10^{-4}$ ($0.0001$)
    \item \textbf{Validation Frequency:} 50 Iterations
\end{itemize}

\subsubsection{Performance Metrics}
The success of the models was analyzed by recording the results obtained on the test dataset into the \texttt{Result\_Metrics.csv} file. Accuracy, Precision, Recall, and F1-Score metrics were used in the evaluation. For the visualization of classification success, Confusion Matrices and \texttt{Result\_TrainingHistory.png} graphs showing the training process were used instead of ROC curves.

\begin{equation}
Accuracy = \frac{TP + TN}{TP + TN + FP + FN}
\end{equation}

\begin{equation}
F1\text{-}Score = 2 \times \frac{Precision \times Recall}{Precision + Recall}
\end{equation}

These metrics are critical for measuring not only the general accuracy of the model but especially its distinctiveness between tumor types.

\label{Sec:eval}
\section{Evaluation of Obtained Data}

In this section, the experimental data obtained from the training and test processes of the proposed custom CNN architecture (OkanNet) and the Transfer Learning-based ResNet-50 model are analyzed in detail. The performance of the models is interpreted in terms of training stability, classification metrics, discrimination power between classes, and computational costs in light of similar studies in the literature. All numerical data were compiled from the \texttt{Result\_Metrics.csv} file automatically created as a result of the experiments.

\subsection{Training Process and Convergence Analysis}
\begin{figure}[htbp]
    \centering
    \resizebox{8.5cm}{!}{\includegraphics{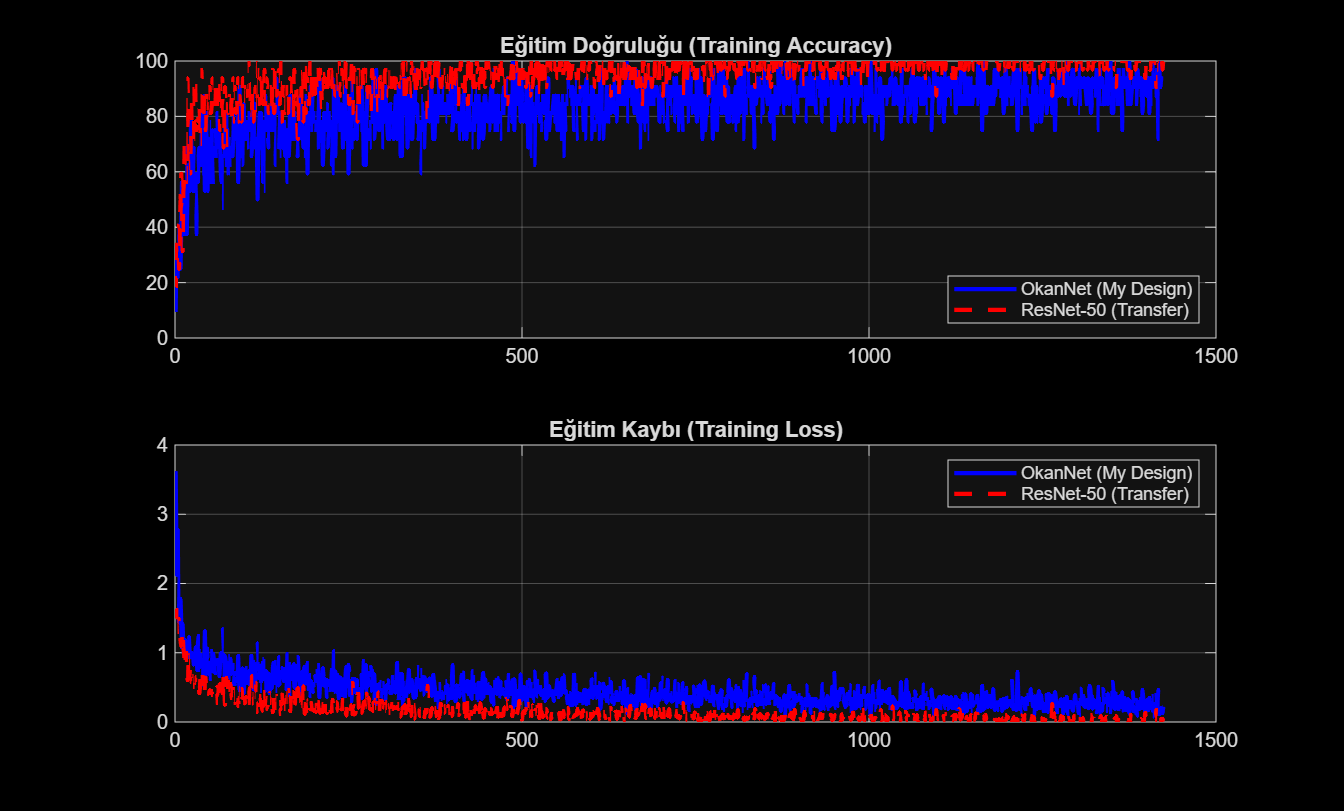}}
    \caption{Training Processes of Models. The blue line (OkanNet) represents learning from scratch, while the red dashed line (ResNet-50) represents the adaptation of transferred knowledge.}
    \label{fig:training_history}
\end{figure}

Both models were trained for $8$ epochs on NVIDIA RTX 2060 GPU hardware. The changes in success (Accuracy) and loss (Loss) functions during the training process with respect to iterations are visualized in Figure \ref{fig:training_history}.

When the graphs are examined, two different learning dynamics are striking:
\begin{enumerate}
    \item \textbf{ResNet-50 (Transfer Learning):} The model shown with the red curve started training from a very high accuracy level ($80\%+$) thanks to millions of parameters transferred from the ImageNet dataset. The model exceeded the $90\%$ band before the first epoch was completed and exhibited rapid convergence.
    \item \textbf{OkanNet (Custom Architecture):} The blue curve reflects the typical learning process of a network initialized with random weights (He Initialization). Starting with low success in the first iterations, the model achieved stable learning at the end of each epoch.
\end{enumerate}

\subsection{Numerical Performance Comparison}
The final achievements of the models on the test dataset (Testing Folder) are presented in Table \ref{tab:results}.

\begin{table}[htbp]
\caption{Performance Metrics of Models}
\begin{center}
\resizebox{8.5cm}{!}{%
\begin{tabular}{|l|c|c|}
\hline
\textbf{Performance Metric} & \textbf{OkanNet} & \textbf{ResNet-50} \\
\hline
\textbf{Accuracy} & \textbf{\%88.10} & \textbf{\%96.49} \\
Precision & 0.877 & 0.963 \\
Recall & 0.872 & 0.962 \\
F1-Score & 0.875 & 0.962 \\
\hline
\textbf{Training Time} & \textbf{311 sec ($\sim$5 min)} & \textbf{1000 sec ($\sim$16 min)} \\
\hline
\end{tabular}%
}
\label{tab:results}
\end{center}
\end{table}

When the data in Table \ref{tab:results} is analyzed:
\begin{itemize}
    \item \textbf{ResNet-50} met the high reliability standards accepted in medical diagnostic systems with $96.49\%$ overall accuracy and $0.962$ F1-Score.
    \item \textbf{OkanNet} exhibited a highly competitive performance despite being a shallow architecture with an $88.10\%$ accuracy rate. The balanced Precision and Recall values indicate that the model is not biased against any class.
\end{itemize}

\subsection{Confusion Matrix Analysis}
\begin{figure}[htbp]
    \centering
    \resizebox{8.5cm}{!}{\includegraphics{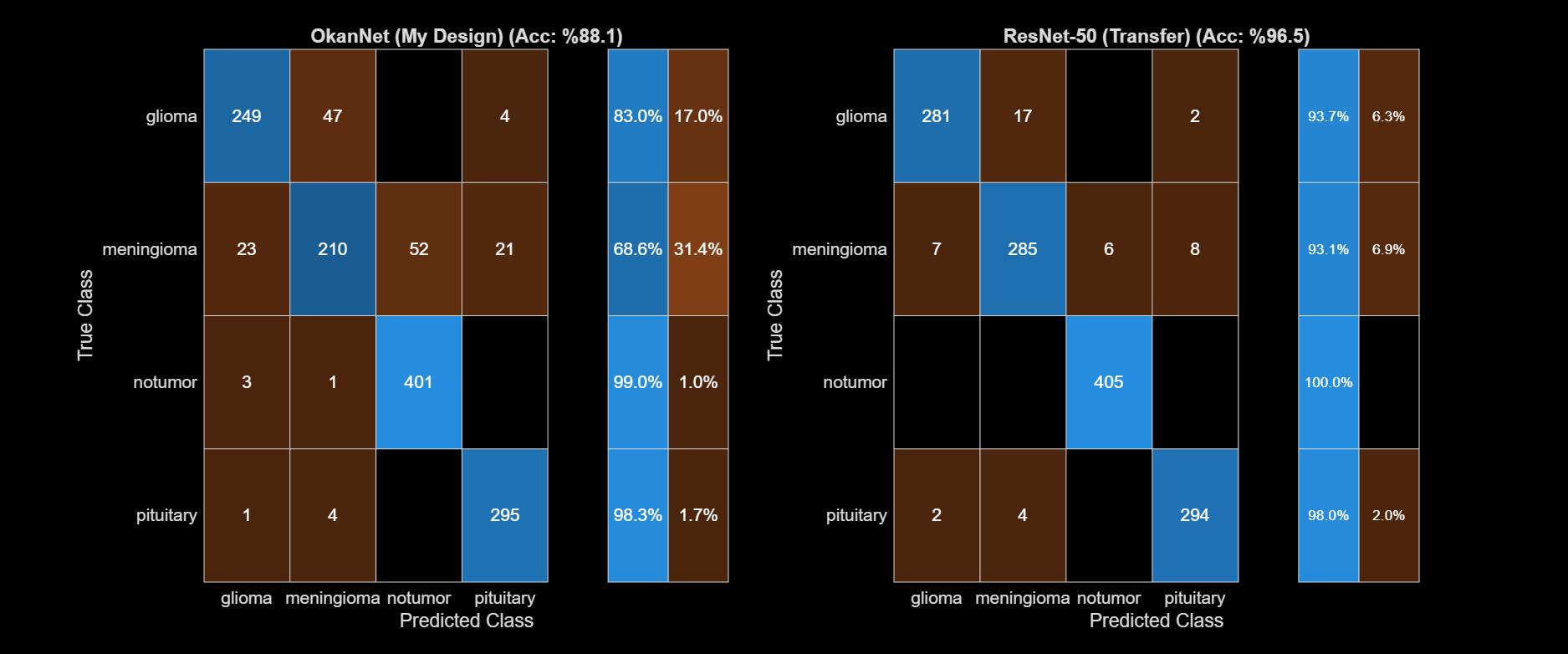}}
    \caption{Confusion Matrices. Left: OkanNet, Right: ResNet-50.}
    \label{fig:confusion}
\end{figure}

Confusion Matrices were examined in Figure \ref{fig:confusion} to understand the root causes of classification errors.

The analysis results show the following:
\begin{itemize}
    \item Both models showed over $98\%$ success in distinguishing the \textbf{"No Tumor"} (Healthy) class.
    \item \textbf{Source of Error:} In the OkanNet model, the vast majority of errors were concentrated between the \textit{Glioma} and \textit{Meningioma} classes. As stated in the medical literature, these two tumor types can show similar tissue characteristics in radiological images.
\end{itemize}

\subsection{Visual Verification and Computational Cost}
\begin{figure}[htbp]
    \centering
    \resizebox{8.5cm}{!}{\includegraphics{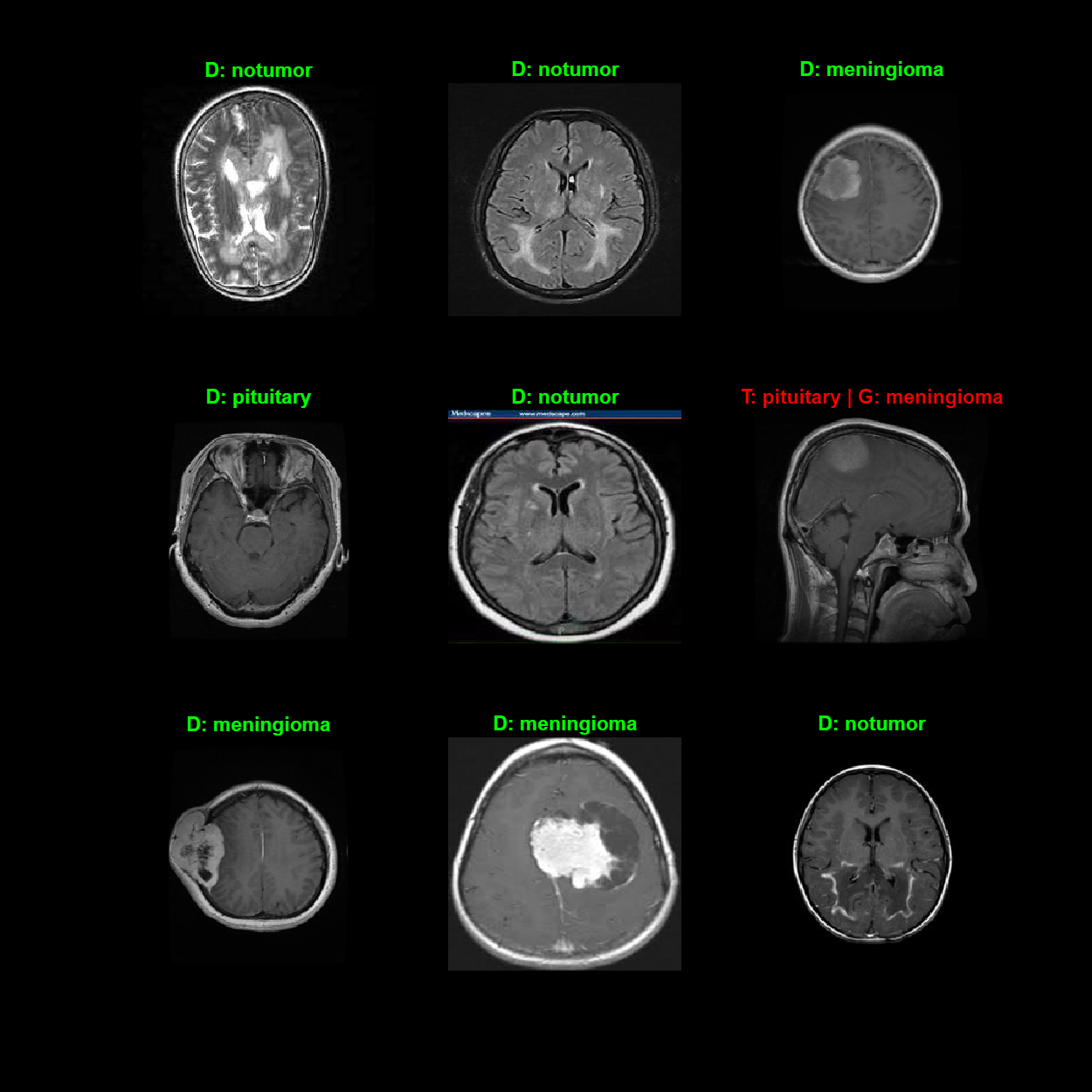}}
    \caption{Sample Predictions Performed with ResNet-50 Model (Green: Correct).}
    \label{fig:predictions}
\end{figure}

The success of the model on real-world data is visualized in Figure \ref{fig:predictions} on samples randomly selected from the test set.

\textbf{Computational Cost and Trade-off:}
The most striking finding of this study is the relationship between performance and time. Although the ResNet-50 model provides $8.3\%$ higher accuracy compared to OkanNet, it is $3.2$ times slower in terms of training time.
\begin{itemize}
    \item \textbf{ResNet-50} should be preferred in servers with high processing power.
    \item \textbf{OkanNet} is more efficient in mobile devices or portable MRI units with the speed and low resource consumption it offers.
\end{itemize}

\label{Sec:conc}
\section{Conclusions and Future Works}
In this study, we presented OkanNet, which is a lightweight deep learning technique for the automatic classification and diagnosis of brain tumors from MRI images. We validated our OkanNet by using the ``Brain Tumor MRI Dataset'' compiled by Masoud Nickparvar~\cite{dataset_ref}. We also compared OkanNet with ResNet-50~\cite{resnet_ref} based transfer learning model on a comprehensive dataset~\cite{dataset_ref} consisting of $7,023$ images. We showed that our OkanNet has acceptable accuracy with low computation times. 

As a result of the experimental findings and analyses obtained, the following judgments were reached:
\begin{enumerate}
    \item \textbf{Transfer Learning Superiority:} The pre-trained ResNet-50 architecture achieved a very high accuracy rate of $96.49\%$ despite being trained with a limited number of epochs ($8$ cycles). This proves that the Transfer Learning method is a strong candidate for clinical decision support systems even if the dataset is limited in medical imaging.
    \item \textbf{Efficiency of Custom Architecture:} The OkanNet architecture designed and trained from scratch exhibited a performance competitive with shallow networks in the literature with an $88.10\%$ success rate. More importantly, OkanNet's training time ($311$ seconds) is approximately \textbf{$3.2$ times faster} compared to the ResNet-50 model ($1000$ seconds).
    \item \textbf{Clinical Applicability:} Error matrix analyses showed that both models work almost flawlessly in distinguishing ``Healthy'' (No Tumor) brain tissue. However, in the distinction of tumor types with similar tissue structures such as Glioma and Meningioma, the sensitivity of ResNet-50, which has deeper layers, is higher.
\end{enumerate}

In conclusion; if the priority is high accuracy and precise diagnosis, \textbf{ResNet-50} should be preferred; if the priority is operability on mobile devices and low processor cost, \textbf{OkanNet} should be preferred. 

In the future, we would like to test more modern hybrid architectures (e.g., EfficientNet) and use synthetic data generation (GANs) methods to eliminate class imbalance. We also would like to implement our OkanNet on mobile platforms. We believe that our fast and accurate deep learning algorithm will work on mobile platforms at real-time frame rates. Additionally, there is potential for upgrading and optimizing our OkanNet by incorporating state of the art deep learning methods and
techniques~\cite{Gok2023SIU, Azadvatan2024arXiv, Akdogan2024arXiv, Jabbarli2024arXiv,Akdogan2025ARTEMIS}.

To improve classification accuracy of our OkanNet, we are also interested in implementing realistic Bidirectional Reflectance Distribution Function (BRDF)~\cite{Ozturk2006EGUK, Kurt2007MScThesis, Ozturk2008CG, Kurt2008SIGGRAPHCG, Kurt2009SIGGRAPHCG, Kurt2010SIGGRAPHCG, Ozturk2010GraphiCon, Szecsi2010SCCG, Ozturk2010CGF, Bigili2011CGF, Bilgili2012SCCG, Ergun2012SCCG, Toral2014SIU, Tongbuasirilai2017ICCVW, Kurt2019DEU, Tongbuasirilai2020TVC, Akleman2024arXiv,Kurt2025arXiv}, Bidirectional Scattering Distribution Function (BSDF)~\cite{WKB12, Ward2014MAM, Kurt2014WLRS, Kurt2016SIGGRAPH, Kurt2017MAM, Kurt2018DEU}, Bidirectional Surface Scattering Reflectance Distribution Function (BSSRDF)~\cite{Kurt2013TPCG, Kurt2013EGSR, Kurt2014PhDThesis, Onel2019PL, Kurt2020MAM, Kurt2021TVC, Yildirim2024arXiv} and multi-layered material~\cite{WKB12, Kurt2016SIGGRAPH, Mir2022DEU} models into our deep learning architecture (OkanNet) as a preprocessing step.

\label{Sec:appendix}
\section{Appendix: MATLAB Source Codes}
Below is the complete MATLAB code used for the realization of the study, covering the entire process (Data processing, Model training, Testing, and Reporting). The codes were developed using MATLAB R2023b and Deep Learning Toolbox.

\lstset{
    language=Matlab,
    basicstyle=\tiny\ttfamily, 
    breaklines=true, 
    keywordstyle=\color{blue},
    commentstyle=\color{green!40!black},
    stringstyle=\color{purple},
    numbers=left,
    numberstyle=\tiny\color{gray},
    frame=single,
    showstringspaces=false,
    captionpos=b,
    title=\lstname
}

\begin{lstlisting}[caption={Brain Tumor Classification Project Complete Source Code}]
% -------------------------------------------------------------------------
% COMPUTER VISION AND DEEP LEARNING - FINAL PROJECT 
% Subject: Brain Tumor Classification (OkanNet vs ResNet-50)
% Prepared By: Okan Ucar
% Hardware: GPU (RTX 2060)
% Features: 
% 1. Automatic Dataset Visualization
% 2. OkanNet Custom Design
% 3. ResNet-50 Transfer Learning
% 4. Visual Prediction Results
% 5. Detailed Reporting (PNG & CSV Export)
% -------------------------------------------------------------------------

clear; clc; close all;

% =========================================================================
% 1. DATASET AND PREPARATION
% =========================================================================
datasetFolder = 'BrainTumorData';
trainFolder = fullfile(datasetFolder, 'Training');
testFolder  = fullfile(datasetFolder, 'Testing');

% Folder check
if ~isfolder(trainFolder)
    error('ERROR: "BrainTumorData" folder not found! Make sure to extract the Zip file to the project folder.');
end

fprintf('Loading dataset... \n');

% Load data
imdsTrain = imageDatastore(trainFolder, 'IncludeSubfolders', true, 'LabelSource', 'foldernames');
imdsTest  = imageDatastore(testFolder, 'IncludeSubfolders', true, 'LabelSource', 'foldernames');

% Class Names
classNames = categories(imdsTrain.Labels);
numClasses = numel(classNames);

disp('Detected Classes:');
disp(classNames);

% -------------------------------------------------------------------------
% FEATURE 1: RANDOM SAMPLES FROM DATASET (DATA PREVIEW)
% -------------------------------------------------------------------------
fprintf('Visualizing samples from dataset...\n');
numSamples = 9;
idx = randperm(numel(imdsTrain.Files), numSamples);
f_preview = figure('Name', 'Dataset Samples', 'Position', [100, 100, 800, 800], 'Visible', 'on');

for i = 1:numSamples
    subplot(3, 3, i);
    I = readimage(imdsTrain, idx(i));
    imshow(I);
    title(char(imdsTrain.Labels(idx(i))), 'FontSize', 12);
end
% Saved the visual to put in the report
saveas(f_preview, 'Result_Dataset_Preview.png'); 

% -------------------------------------------------------------------------
% DATA PREPROCESSING CONTINUED
% -------------------------------------------------------------------------
inputSize = [224 224 3]; % ResNet standard

% Resize & Grayscale to RGB
augimdsTrain = augmentedImageDatastore(inputSize(1:2), imdsTrain, 'ColorPreprocessing', 'gray2rgb');
augimdsTest  = augmentedImageDatastore(inputSize(1:2), imdsTest, 'ColorPreprocessing', 'gray2rgb');

% Training Settings
options = trainingOptions('sgdm', ...
    'MiniBatchSize', 32, ...
    'MaxEpochs', 8, ...             
    'InitialLearnRate', 1e-4, ...
    'Shuffle', 'every-epoch', ...
    'ValidationData', augimdsTest, ...
    'ValidationFrequency', 50, ...
    'Verbose', false, ...
    'Plots', 'training-progress', ... 
    'ExecutionEnvironment', 'auto'); 

% =========================================================================
% MODEL 1: CUSTOM CNN (OkanNet)
% =========================================================================
myModelName = 'OkanNet (My Design)'; 
fprintf('\n--- MODEL 1: %s Training... ---\n', myModelName);

layersCustom = [
    imageInputLayer(inputSize, 'Name', 'Input')
    
    % Block 1
    convolution2dLayer(3, 16, 'Padding', 'same', 'Name', 'Conv_1')
    batchNormalizationLayer('Name', 'BN_1')
    reluLayer('Name', 'ReLU_1')
    maxPooling2dLayer(2, 'Stride', 2, 'Name', 'Pool_1')
    
    % Block 2
    convolution2dLayer(3, 32, 'Padding', 'same', 'Name', 'Conv_2')
    batchNormalizationLayer('Name', 'BN_2')
    reluLayer('Name', 'ReLU_2')
    maxPooling2dLayer(2, 'Stride', 2, 'Name', 'Pool_2')
    
    % Block 3
    convolution2dLayer(3, 64, 'Padding', 'same', 'Name', 'Conv_3')
    batchNormalizationLayer('Name', 'BN_3')
    reluLayer('Name', 'ReLU_3')
    maxPooling2dLayer(2, 'Stride', 2, 'Name', 'Pool_3')
    
    % Classification
    fullyConnectedLayer(128, 'Name', 'FC_1')
    reluLayer('Name', 'ReLU_FC')
    dropoutLayer(0.5, 'Name', 'Dropout')
    fullyConnectedLayer(numClasses, 'Name', 'FC_Out')
    softmaxLayer('Name', 'Softmax')
    classificationLayer('Name', 'Output')
];

tic;
[netCustom, infoCustom] = trainNetwork(augimdsTrain, layersCustom, options);
timeCustom = toc;
fprintf('%s Completed. Time: %.2f sec\n', myModelName, timeCustom);

% =========================================================================
% MODEL 2: TRANSFER LEARNING (ResNet-50)
% =========================================================================
refModelName = 'ResNet-50 (Transfer)';
fprintf('\n--- MODEL 2: %s Training... ---\n', refModelName);

try
    netRes = resnet50;
catch
    error('ResNet-50 package missing! Install from Add-Ons.');
end

lgraph = layerGraph(netRes);
newFCLayer = fullyConnectedLayer(numClasses, 'Name', 'NewFC', 'WeightLearnRateFactor', 10, 'BiasLearnRateFactor', 10);
newClassLayer = classificationLayer('Name', 'NewOutput');
lgraph = replaceLayer(lgraph, 'fc1000', newFCLayer);
lgraph = replaceLayer(lgraph, 'ClassificationLayer_fc1000', newClassLayer);

tic;
[netTransfer, infoTransfer] = trainNetwork(augimdsTrain, lgraph, options);
timeTransfer = toc;
fprintf('%s Completed. Time: %.2f sec\n', refModelName, timeTransfer);

% =========================================================================
% CALCULATION OF RESULTS
% =========================================================================
fprintf('\n--- TEST AND REPORTING ---\n');

% Predictions
[YPred1, scores1] = classify(netCustom, augimdsTest);
accCustom = mean(YPred1 == imdsTest.Labels) * 100;

[YPred2, scores2] = classify(netTransfer, augimdsTest);
accTransfer = mean(YPred2 == imdsTest.Labels) * 100;

% =========================================================================
% VISUALIZATION AND SAVING
% =========================================================================

% 1. Confusion Matrix
f1 = figure('Name', 'Final Comparison: Confusion Matrix', 'Position', [100, 100, 1200, 500]);
subplot(1, 2, 1);
cm1 = confusionchart(imdsTest.Labels, YPred1);
cm1.Title = [myModelName ' (Acc: %' num2str(accCustom, '%.1f') ')'];
cm1.RowSummary = 'row-normalized'; 
subplot(1, 2, 2);
cm2 = confusionchart(imdsTest.Labels, YPred2);
cm2.Title = [refModelName ' (Acc: %' num2str(accTransfer, '%.1f') ')'];
cm2.RowSummary = 'row-normalized'; 
saveas(f1, 'Result_ConfusionMatrix.png');

% 2. Training Graph
f2 = figure('Name', 'Training Performance Comparison');
subplot(2,1,1);
plot(infoCustom.TrainingAccuracy, 'LineWidth', 2, 'Color', 'b'); hold on;
plot(infoTransfer.TrainingAccuracy, 'LineWidth', 2, 'Color', 'r', 'LineStyle', '--');
legend(myModelName, refModelName, 'Location', 'southeast');
title('Training Accuracy'); grid on;
subplot(2,1,2);
plot(infoCustom.TrainingLoss, 'LineWidth', 2, 'Color', 'b'); hold on;
plot(infoTransfer.TrainingLoss, 'LineWidth', 2, 'Color', 'r', 'LineStyle', '--');
legend(myModelName, refModelName);
title('Training Loss'); grid on;
saveas(f2, 'Result_TrainingHistory.png');

% -------------------------------------------------------------------------
% FEATURE 2: VISUALIZE PREDICTION RESULTS (VISUAL PREDICTIONS)
% -------------------------------------------------------------------------
% Test and show 9 random test images with the best model (ResNet)
fprintf('Visualizing model predictions...\n');
idxTest = randperm(numel(imdsTest.Files), 9);
f3 = figure('Name', 'Model Prediction Samples', 'Position', [100, 100, 800, 800]);

for i = 1:9
    subplot(3, 3, i);
    I = readimage(imdsTest, idxTest(i));
    
    % Resize image (for display)
    I_show = imresize(I, [224 224]);
    imshow(I_show);
    
    % Prediction and True Label
    predLabel = YPred2(idxTest(i)); % ResNet predictions
    trueLabel = imdsTest.Labels(idxTest(i));
    
    % Color Setting (Green if True, Red if False)
    if predLabel == trueLabel
        titleColor = 'g'; % Green
        titleText = sprintf('P: %s', char(predLabel));
    else
        titleColor = 'r'; % Red
        titleText = sprintf('P: %s | T: %s', char(predLabel), char(trueLabel));
    end
    
    title(titleText, 'Color', titleColor, 'FontSize', 11, 'FontWeight', 'bold');
end
% Save
saveas(f3, 'Result_Prediction_Visuals.png');

% =========================================================================
% DETAILED METRICS AND CSV
% =========================================================================
fprintf('\n=======================================================\n');
fprintf('           DETAILED PERFORMANCE REPORT                 \n');
fprintf('=======================================================\n');

% Metric Calculation
confMat1 = confusionmat(imdsTest.Labels, YPred1);
prec1 = mean(diag(confMat1) ./ sum(confMat1, 1)', 'omitnan'); 
rec1  = mean(diag(confMat1) ./ sum(confMat1, 2), 'omitnan');  
f1_1  = 2 * (prec1 * rec1) / (prec1 + rec1);                  

confMat2 = confusionmat(imdsTest.Labels, YPred2);
prec2 = mean(diag(confMat2) ./ sum(confMat2, 1)', 'omitnan');
rec2  = mean(diag(confMat2) ./ sum(confMat2, 2), 'omitnan');
f1_2  = 2 * (prec2 * rec2) / (prec2 + rec2);

fprintf('%-20s | %-12s | %-12s \n', 'METRIC', 'OkanNet', 'ResNet-50');
fprintf('---------------------|--------------|--------------\n');
fprintf('%-20s | %.4f       | %.4f \n', 'Accuracy', accCustom/100, accTransfer/100);
fprintf('%-20s | %.4f       | %.4f \n', 'Precision', prec1, prec2);
fprintf('%-20s | %.4f       | %.4f \n', 'Recall', rec1, rec2);
fprintf('%-20s | %.4f       | %.4f \n', 'F1-Score', f1_1, f1_2);
fprintf('%-20s | %.2f sec     | %.2f sec \n', 'Training Time', timeCustom, timeTransfer);
fprintf('-------------------------------------------------------\n');

metricsTable = table(...
    {'Accuracy'; 'Precision'; 'Recall'; 'F1-Score'; 'Training Time'}, ...
    [accCustom/100; prec1; rec1; f1_1; timeCustom], ...
    [accTransfer/100; prec2; rec2; f1_2; timeTransfer], ...
    'VariableNames', {'Metric', 'OkanNet', 'ResNet50'});

writetable(metricsTable, 'Result_Metrics.csv');

disp('    ALL OPERATIONS COMPLETED!');
disp('--> Result_Dataset_Preview.png (Data Samples)');
disp('--> Result_ConfusionMatrix.png (Confusion Matrix)');
disp('--> Result_TrainingHistory.png (Training Graph)');
disp('--> Result_Prediction_Visuals.png (Prediction Samples)');
disp('--> Result_Metrics.csv (Numerical Data)');
\end{lstlisting}

\bibliographystyle{IEEEtran}
\bibliography{arXiv26_OkanNet_References}






\end{document}